\documentclass[a4paper,USenglish,cleveref, autoref, thm-restate]{lipics-v2021}

\title{Database Theory in Action: Yannakakis' Algorithm} 

\author{Paraschos Koutris}{University of Wisconsin-Madison, USA}{paris@cs.wisc.edu}{https://orcid.org/0000-0001-6309-1702}{}

\author{Stijn Vansummeren}{UHasselt, Data Science Institute, Belgium}{stijn.vansummeren@uhasselt.be}{https://orcid.org/0000-0001-7793-9049}{}

\author{Qichen Wang}{Nanyang Technological University, Singapore}{qichen.wang@ntu.edu.sg}{https://orcid.org/0000-0002-0959-5536}{}

\author{Yisu Remy Wang}{University of California, Los Angeles, USA}{remywang@cs.ucla.edu}{https://orcid.org/0000-0002-6887-9395}{}

\author{Xiangyao Yu}{University of Wisconsin-Madison, USA}{yxy@cs.wisc.edu}{https://orcid.org/0009-0001-0785-2519}{}

\authorrunning{P. Koutris, S. Vansummeren, Q. Wang, Y.\,R. Wang, and X. Yu}

\Copyright{Paraschos Koutris, Stijn Vansummeren, Qichen Wang, Yisu Remy Wang, and Xiangyao Yu}



\relatedversion{} 

\nolinenumbers 

\usepackage{tikz}

\usetikzlibrary{positioning}
\usetikzlibrary{trees}
\usetikzlibrary{calc}
\usetikzlibrary{tikzmark}
\usetikzlibrary{matrix}

\newcommand{\YA}{\textsf{YA}}
\newcommand{\PT}{\textsf{PT}}
\newcommand{\RPT}{\textsf{RPT}}
\newcommand{\Yplus}{\textsf{YA$^+$}}
\newcommand{\setof}[2]{\left\{#1 \mid #2\right\}}

\newcommand{\nsemijoinsymb}{\tikz[baseline=-2pt]{\draw (0,-0.6ex)--(0.5ex,0)--(0,0.6ex)--(0,-0.6ex);\draw[->, >=stealth,draw] (0.5ex,0)--(1.75ex,0);}}

\DeclareMathOperator{\nsemijoin}{\nsemijoinsymb}

\EventEditors{Balder ten Cate and Maurice Funk}
\EventNoEds{2}
\EventLongTitle{29th International Conference on Database Theory (ICDT 2026)}
\EventShortTitle{ICDT 2026}
\EventAcronym{ICDT}
\EventYear{2026}
\EventDate{March 24--27, 2026}
\EventLocation{Tampere, Finland}
\EventLogo{}
\SeriesVolume{365}
\ArticleNo{21}
\category{Database Theory in Action}

\begin{document}

\maketitle

\begin{abstract}
Yannakakis' seminal algorithm is optimal for acyclic joins, 
yet it has not been widely adopted due to its
poor performance in practice. 
This paper briefly surveys recent advancements in
making Yannakakis' algorithm more practical, 
in terms of both efficiency and ease of implementation,
and points out several avenues for future research.
\end{abstract}

\section{Introduction}\label{sec:intro}
In 1981, an optimal algorithm for computing acyclic joins was concurrently discovered by Bernstein et al.~\cite{DBLP:journals/jacm/BernsteinC81, DBLP:journals/tods/BernsteinGWRR81} and Yannakakis~\cite{DBLP:conf/vldb/Yannakakis81}, and has since become known as Yannakakis' algorithm (hereafter $\YA$). 
The algorithm is conceptually simple, applicable to a wide class of practical queries, and has
the strong guarantee of being {\em instance optimal}, which means it has the
best possible asymptotic complexity for {\em any} input instance.  
Concretely, recall that a natural join query
$Q = R_1(\mathbf{x}_1) \Join \dots \Join  R_n(\mathbf{x}_n)$ is {\em
  $\alpha$-acyclic} if there exists a {\em join tree}, which is a tree whose
nodes are the relations of $Q$, and for each variable $x$ in $Q$, the set of
nodes $\setof{R_i}{x \in \mathbf{x}_i}$ is connected. To evaluate an $\alpha$-acyclic query $Q$,
 $\YA$ makes two passes over the query's join tree,
 using semijoins to remove dangling tuples;
 then it makes a final pass to produce the output.
For example, given $Q = R(i, j) \Join S(j, k) \Join T(k, l) \Join U(l, m)$ with join tree%
\footnote{Each edge points from parent to child.}
$R \rightarrow S \rightarrow T \rightarrow U$, 
$\YA$ computes the following relations in sequence: 
$T' = T\ltimes U$, $S'  = S \ltimes T'$, $R^*  =R\ltimes T'$ (bottom-up pass); 
$S^*  = S' \ltimes R^*$, $ T^*  = T'\ltimes S^*$, $U^*  = U  \ltimes T^*$ (top-down pass); and 
$Q_1  = T^* \Join U^*$, $Q_2  = S^*\Join Q_1$, $Q  = R^* \Join Q_2$ (final output).

Despite $\YA$'s simplicity and instance-optimality, none of
today's mainstream database systems implement the algorithm, mainly due to the
high constant factor overhead involved in performing the semijoin reduction phases. 
For instance, recent experiments by Gottlob et al.~\cite{gottlob2023structure} demonstrate that while $\YA$ improves tail-end performance for long-running queries, it increases run time by an average of $2.4\times$ compared to standard binary joins. Furthermore, the algorithm's three-pass design 
complicates its integration into existing query optimizers. 
 In response to these observations, recent research has proposed
several approaches to make $\YA$ more practical: more efficient
semijoin operator implementations; reducing the number of semijoin passes;  and
creating novel join operators that can achieve the theoretical benefits of $\YA$
without additional overhead. 
In this paper we provide a brief survey of these approaches, distill the key
insights behind them, and identify open challenges for future research.

\section{Improving Semijoin Efficiency with Bitvector Filters}\label{sec:filters}

Bloom filter~\cite{DBLP:journals/cacm/Bloom70} 
is a space-efficient, probabilistic data structure for sets. By mapping each element in a set to $k$ bit positions in a vector using $k$ independent hash functions, it supports set membership queries with a configurable one-sided error: 
false positives are possible but there will be
no false negatives.   
This property makes Bloom filters very effective for pre-filtering join relations while lowering memory usage, and they are more cache-friendly compared to  hash tables.
To perform a pre-filtering pass for a binary join $R \Join S$, a system can first build a Bloom filter on the join keys from one relation (e.g., $S$) and then probe this filter with the join keys from the other relation ($R$). Tuples in $R$ whose keys fail the membership test are guaranteed not to join with $S$ and can be safely discarded. While false positives may allow some non-joining tuples to pass through, these are subsequently eliminated during the final join on the pre-filtered relations.  

Integrating bitvector filters such as Bloom filters  directly into a cost-based optimizer is non-trivial, as they can cause the plan space to grow exponentially with the number of relations. However, for the common cases of star and snowflake schemas with primary-key-foreign-key joins, it has been shown that the optimal plan within the space of right-deep trees can be found by evaluating only a linear number of candidate plans, which are robust against the join ordering of dimension tables \cite{bitvector2020Ding}. Further research has focused on integrating the selection of filter types \cite{Lang2019} and their configuration into the optimizer's cost model \cite{BloomSIGMOD25}. 

Recently, {\em predicate transfer} ($\PT$)~\cite{DBLP:conf/cidr/YangZYK24} was proposed to approximate 
 semijoin 
using Bloom filters.  
By decoupling the pre-filtering phase from the join phase, $\PT$ improves the robustness of the subsequent join ordering decision. Using a simple heuristic of transferring predicates from smaller relations to larger ones, $\PT$ often achieves similar performance across different join orders.  Building on this, {\em Robust Predicate Transfer} ($\RPT$)~\cite{Debuking2025Zhao} refines $\PT$ to restore the theoretical guarantees of $\YA$ for acyclic queries and implement it into DuckDB's query engine.  Instead of the simple small-to-large assumption, $\RPT$ chooses a ``root'' relation to maximize the downstream reduction.  Furthermore, $\RPT$ requests the subsequent join order to be {\em monotone}.  When the query is $\gamma$-acyclic, any join order is monotone, while for general $\alpha$-acyclic queries, such monotone orders will correspond to a specific join tree.

Several research challenges remain. Integrating bitvector filters into a cost-based optimizer is essential, as the overhead of the pre-filtering phase may not be justified for unselective queries. Further investigation is needed to better understand how to leverage existing indexes, such B-trees. Finally, joins and filters often come with downstream operations like aggregations.  When the bitvector filter needs to carry payloads, one can no longer benefit from the Bloom filter, as the false positives can violate the final correctness of the query.

\section{Unlocking Theoretical Efficiency with Query Rewriting}\label{sec:rewrite}

$\YA$ is a pure {\em relational} algorithm \cite{relational}; its logic is data independent and can be expressed as a 
composition of  entirely 
standard relational operators. This property makes it possible to implement $\YA$-style evaluation via query rewriting \cite{gottlob2023structure, Yannakakis+, Aggregate2025VLDB}, which separates planning from execution. An external planner first constructs a structure-aware query plan based on $\YA$, which is then translated into SQL for execution by an unmodified DBMS. This translation can take the form of a script of sequential SQL statements that materialize intermediate subqueries \cite{gottlob2023structure}, or a single, complex DAG plan that allows the underlying engine to perform more holistic optimizations such as operator pipelining \cite{Yannakakis+}.

Nevertheless, a rewrite-based implementation of $\YA$ can incur performance overhead, particularly on queries with simple PK-FK joins where traditional optimizers are highly effective \cite{gottlob2023structure, Yannakakis+}. Recent work has focused on algorithmic refinements of $\YA$ itself to minimize the number of semijoins required. In fact, the three-phase $\YA$ can collapse into two phases by eliminating the top-down semijoins, when the join order respects the hierarchical structure of the join tree \cite{DBLP:conf/csl/BaganDG07}.  Building on this insight, $\Yplus$ \cite{Yannakakis+} formalizes a cost-based optimization for $\YA$. The high-level idea is to maximize the filtering power of the single semijoin round by carefully choosing the join tree and the semijoin order. For an acyclic query, $\Yplus$ enumerates all join trees for the query and estimates the size of intermediate results after a bottom-up semijoin pass for each tree. It then selects the join tree and corresponding semijoin scheduling that minimizes the intermediate result sizes. This strategy yields much more robust performance compared to conventional binary join plans: it incurs virtually no slowdown in the worst case, yet it can still achieve order-of-magnitude speedups on long-running queries.

Extending $\YA$-style evaluation to other relational operators like aggregation is also realized through query rewriting. For example, aggregation can be supported by evaluating $\YA$ over annotated relations, where each tuple carries annotations used to compute aggregates \cite{AJAR2016}. This abstraction allows aggregation to be pushed down. $\Yplus$ \cite{Yannakakis+} integrates this idea with its two-phase strategy, maintaining annotations during the semijoins. Furthermore, when a query’s group-by attributes are entirely contained in a single relation, the query is said to be relation-dominated \cite{Yannakakis+} or 0MA \cite{Aggregate2025VLDB}. In such cases, the final join phase of $\YA$ can be omitted, and aggregation can be performed directly on the filtered relation after semijoin pruning.

Query rewriting offers a lightweight and portable path to integrate $\YA$-style evaluation into existing database systems without modifying their core components. It is also orthogonal to physical-level optimizations, such as using Bloom filters.  However, this approach introduces certain overheads in practice. Because the logical plan is fixed externally, the rewritten queries may bypass some of the system’s built-in optimization rules, potentially leading to suboptimal plans. The workflow also incurs extra planning time and communication costs during optimization. While the rewriting approach is effective for proof-of-concept deployments, it motivates the need for native support of $\YA$-style strategies within the optimizer and execution engine. A direct implementation could preserve the structural advantages of $\YA$ while minimizing the operational overhead of external rewrites.




\section{Achieving Optimality without Regression}
The techniques discussed so far {\em reduce} the overhead of $\YA$ in different ways.
In this section, we turn to a class of recent algorithms that have provable
 guarantees to incur {\em zero overhead}.
%
An algorithm $A$ is said to be zero-overhead with respect to a baseline algorithm $B$ if the former has lower cost (according to some cost model) for every query plan.
Specifically, the goal is to design algorithms
that are zero-overhead with respect to the binary hash join.

Traditional hash join consists of two phases: the first phase builds hash tables
 for certain relations, and the second probes into these hash tables
 to find matching tuples.
The intuition behind zero-overhead algorithms is to ``piggyback'' semijoin reduction
 into one of these phases, thereby achieving instance-optimality ``for free''.
The algorithms fall into two classes:
 bottom-up algorithms perform semijion reduction during the build phase,
 while top-down algorithms do so during the probe phase.
These classes are dual to each other: while bottom-up algorithms guarantee zero overhead
 for right-deep query plans, top-down algorithms assume left-deep plans.\footnote{We adopt the
 convention to build hash tables on the right side of each binary join operator.}
Bottom-up algorithms are therefore most effective when hash table construction accounts for
 the main cost during execution,
 whereas top-down algorithms reduce the cost of probing hash tables
 and perform well when indexes are available.

The first bottom-up algorithm was proposed by Birler, Kemper, and Neumann~\cite{DBLP:journals/pvldb/BirlerKN24}
 for compiled query engines,
 and later formalized using nested relational algebra by Bekkers et al.~\cite{DBLP:journals/pvldb/BekkersNVW25} who
 generalized it to support other relational operators
 and adapted it for vectorized, column-oriented systems.
The key idea of these algorithms is to introduce a new
 {\em nested semijoin} operator $\nsemijoin$ that is ``in between''
 a join and a semijoin.
During the execution of $R \nsemijoin S$,
 the operator scans $R$ and probes into the hash table for $S$.
For each tuple $r\in R$, instead of materializing all matching tuples
 $s \in S$, the operator attaches to $r$ a pointer to these matches,
 thereby {\em nesting} the matches under $r$.
The algorithms then build up a nested representation of the join result
 by computing the nested semijoins bottom-up along the join tree,
 and finally unnest the representation to produce the output.
Throughout this process, each ``nested join'' runs in time
 linear in the size of each input relation,
 and the final unnesting step takes linear time in the size
 of the output.
Bekkers et al.\ proved the algorithm to be zero-overhead
 for a class of plans that generalizes right-deep plans.

%
%
%
%

Hu, Wang, and Miranker~\cite{hu2024treetracker} proposed a top-down algorithm called TreeTracker Join.
They present the algorithm as a simple modification of the traditional (pipelined)
 binary hash join: whenever a hash lookup fails, 
 the algorithm backtracks to the tuple responsible for the failure
 and removes it from its relation.
To demonstrate the main ideas of the algorithm, let us consider the same query $Q$
 in~\cref{sec:intro}.
Given the left-deep query plan $(R \Join (S \Join (T \Join U)))$,
 and assume hash tables are built for $S$, $T$, and $U$,
 the algorithm first iterates over $R$, probing into the hash table for $S$.
Suppose for the tuple $(i_1, j_1) \in R$ this results in a match $(j_1, k_1) \in S$
 which the algorithm uses to probe into $T$.
This again returns a match $(k_1, l_1) \in T$, but probing into
 $U$ with that tuple fails to return any match.
Because the join key $l_1$ was introduced by the $T$ relation, the algorithm proceeds
 to delete $(k_1, l_1)$ from $T$,
 thereby removing that dangling tuple ``on the fly''.
Hu, Wang, and Miranker proved the algorithm to be instance-optimal
 and zero-overhead for left-deep plans.

Compared to other approaches, 
 implementing the new operators in zero-overhead algorithms requires greater effort.
On the other hand, these algorithms offer real performance gains,
 and their strong guarantee of regression avoidance is critical
 for practical deployment.

\section{Conclusion and Open Problems}
Each approach to improve $\YA$ discussed makes a different tradeoff
between ease of implementation and performance, 
and reduces algorithm overhead in different ways. 
To fully realize the potential of optimal join algorithms, 
 however, requires overcoming additional challenges.
For example, although some of the approaches are compatible with indexes,
 fully leveraging all available indexes remains an interesting problem.
There is also ongoing debate on the role of query optimization
 in the context of optimal algorithms: 
 while $\RPT$ has been shown to perform well regardless of
 the query plan, 
 a better plan does lead to faster execution in other approaches.
Query optimization for acyclic joins also requires different methods of plan enumeration and cost estimation.

\bibliography{references}

@article{DBLP:journals/tods/BernsteinGWRR81,
  author       = {Philip A. Bernstein and
                  Nathan Goodman and
                  Eugene Wong and
                  Christopher L. Reeve and
                  James B. Rothnie Jr.},
  title        = {Query Processing in a System for Distributed Databases {(SDD-1)}},
  journal      = {{ACM} Trans. Database Syst.},
  volume       = {6},
  number       = {4},
  pages        = {602--625},
  year         = {1981},
  url          = {https://doi.org/10.1145/319628.319650},
  doi          = {10.1145/319628.319650},
  bibsource    = {dblp computer science bibliography, https://dblp.org}
}

@article{DBLP:journals/jacm/BernsteinC81,
  author       = {Philip A. Bernstein and
                  Dah{-}Ming W. Chiu},
  title        = {Using Semi-Joins to Solve Relational Queries},
  journal      = {J. {ACM}},
  volume       = {28},
  number       = {1},
  pages        = {25--40},
  year         = {1981},
  url          = {https://doi.org/10.1145/322234.322238},
  doi          = {10.1145/322234.322238},
  bibsource    = {dblp computer science bibliography, https://dblp.org}
}

@article{DBLP:journals/cacm/Bloom70,
  author       = {Burton H. Bloom},
  title        = {Space/Time Trade-offs in Hash Coding with Allowable Errors},
  journal      = {Commun. {ACM}},
  volume       = {13},
  number       = {7},
  pages        = {422--426},
  year         = {1970},
  url          = {https://doi.org/10.1145/362686.362692},
  doi          = {10.1145/362686.362692},
  bibsource    = {dblp computer science bibliography, https://dblp.org}
}

@inproceedings{DBLP:conf/vldb/Yannakakis81,
  author       = {Mihalis Yannakakis},
  title        = {Algorithms for Acyclic Database Schemes},
  booktitle    = {Very Large Data Bases, 7th International Conference, September 9-11,
                  1981, Cannes, France, Proceedings},
  pages        = {82--94},
  publisher    = {{IEEE} Computer Society},
  year         = {1981},
  bibsource    = {dblp computer science bibliography, https://dblp.org}
}

@article{DBLP:journals/pvldb/BirlerKN24,
  author       = {Altan Birler and
                  Alfons Kemper and
                  Thomas Neumann},
  title        = {Robust Join Processing with Diamond Hardened Joins},
  journal      = {Proc. {VLDB} Endow.},
  volume       = {17},
  number       = {11},
  pages        = {3215--3228},
  year         = {2024},
  url          = {https://www.vldb.org/pvldb/vol17/p3215-birler.pdf},
  doi          = {10.14778/3681954.3681995},
  bibsource    = {dblp computer science bibliography, https://dblp.org}
}

@article{DBLP:journals/pvldb/BekkersNVW25,
  author       = {Liese Bekkers and
                  Frank Neven and
                  Stijn Vansummeren and
                  Yisu Remy Wang},
  title        = {Instance-Optimal Acyclic Join Processing Without Regret: Engineering
                  the Yannakakis Algorithm in Column Stores},
  journal      = {Proc. {VLDB} Endow.},
  volume       = {18},
  number       = {8},
  pages        = {2413--2426},
  year         = {2025},
  url          = {https://www.vldb.org/pvldb/vol18/p2413-vansummeren.pdf},
  bibsource    = {dblp computer science bibliography, https://dblp.org}
}

@article{hu2024treetracker,
  title={TreeTracker Join: Simple, Optimal, Fast},
  author={Hu, Zeyuan and Wang, Yisu Remy and Miranker, Daniel P},
  journal={arXiv preprint arXiv:2403.01631},
  year={2024}
}

@article{gottlob2023structure,
  title={Structure-guided query evaluation: Towards bridging the gap from theory to practice},
  author={Gottlob, Georg and Lanzinger, Matthias and Longo, Davide Mario and Okulmus, Cem and Pichler, Reinhard and Selzer, Alexander},
  journal={arXiv preprint arXiv:2303.02723},
  year={2023}
}

@article{Yannakakis+,
author = {Wang, Qichen and Chen, Bingnan and Dai, Binyang and Yi, Ke and Li, Feifei and Lin, Liang},
title = {Yannakakis+: Practical Acyclic Query Evaluation with Theoretical Guarantees},
year = {2025},
issue_date = {June 2025},
publisher = {Association for Computing Machinery},
address = {New York, NY, USA},
volume = {3},
number = {3},
url = {https://doi.org/10.1145/3725423},
doi = {10.1145/3725423},
journal = {Proc. ACM Manag. Data},
month = jun,
articleno = {235},
numpages = {28}
}

@article{relational,
author = {Wang, Qichen and Luo, Qiyao and Wang, Yilei},
title = {Relational Algorithms for Top-k Query Evaluation},
year = {2024},
issue_date = {June 2024},
publisher = {Association for Computing Machinery},
address = {New York, NY, USA},
volume = {2},
number = {3},
url = {https://doi.org/10.1145/3654971},
doi = {10.1145/3654971},
journal = {Proc. ACM Manag. Data},
month = may,
articleno = {168},
numpages = {27}
}

@inproceedings{bitvector2020Ding,
author = {Ding, Bailu and Chaudhuri, Surajit and Narasayya, Vivek},
title = {Bitvector-aware Query Optimization for Decision Support Queries},
year = {2020},
isbn = {9781450367356},
publisher = {Association for Computing Machinery},
address = {New York, NY, USA},
url = {https://doi.org/10.1145/3318464.3389769},
doi = {10.1145/3318464.3389769},
booktitle = {Proceedings of the 2020 ACM SIGMOD International Conference on Management of Data},
pages = {2011–2026},
numpages = {16},
location = {Portland, OR, USA},
series = {SIGMOD '20}
}

@article{Lang2019,
author = {Lang, Harald and Neumann, Thomas and Kemper, Alfons and Boncz, Peter},
title = {Performance-optimal filtering: Bloom overtakes Cuckoo at high throughput},
year = {2019},
issue_date = {January 2019},
publisher = {VLDB Endowment},
volume = {12},
number = {5},
issn = {2150-8097},
url = {https://doi.org/10.14778/3303753.3303757},
doi = {10.14778/3303753.3303757},
journal = {Proc. VLDB Endow.},
month = jan,
pages = {502–515},
numpages = {14}
}

@article{Debuking2025Zhao,
author = {Zhao, Junyi and Su, Kai and Yang, Yifei and Yu, Xiangyao and Koutris, Paraschos and Zhang, Huanchen},
title = {Debunking the Myth of Join Ordering: Toward Robust SQL Analytics},
year = {2025},
issue_date = {June 2025},
publisher = {Association for Computing Machinery},
address = {New York, NY, USA},
volume = {3},
number = {3},
url = {https://doi.org/10.1145/3725283},
doi = {10.1145/3725283},
journal = {Proc. ACM Manag. Data},
month = jun,
articleno = {146},
numpages = {28}
}

@inproceedings{BloomSIGMOD25,
author = {Zeyl, Tim and Cheng, Qi and Pournaghi, Reza and Lam, Jason and Wang, Weicheng and Wong, Calvin and Chen, Chong and Larson, Per-Ake},
title = {Including Bloom Filters in Bottom-up Optimization},
year = {2025},
isbn = {9798400715648},
publisher = {Association for Computing Machinery},
address = {New York, NY, USA},
url = {https://doi.org/10.1145/3722212.3724440},
doi = {10.1145/3722212.3724440},
booktitle = {Companion of the 2025 International Conference on Management of Data},
pages = {703–715},
numpages = {13},
location = {Berlin, Germany},
series = {SIGMOD/PODS '25}
}

@inproceedings{DBLP:conf/cidr/YangZYK24,
  author       = {Yifei Yang and
                  Hangdong Zhao and
                  Xiangyao Yu and
                  Paraschos Koutris},
  title        = {Predicate Transfer: Efficient Pre-Filtering on Multi-Join Queries},
  booktitle    = {14th Conference on Innovative Data Systems Research, {CIDR} 2024,
                  Chaminade, HI, USA, January 14-17, 2024},
  publisher    = {www.cidrdb.org},
  year         = {2024},
  url          = {https://www.cidrdb.org/cidr2024/papers/p22-yang.pdf},
  bibsource    = {dblp computer science bibliography, https://dblp.org}
}

@article{Aggregate2025VLDB,
author = {Lanzinger, Matthias and Pichler, Reinhard and Selzer, Alexander},
title = {Avoiding Materialisation for Guarded Aggregate Queries},
year = {2025},
issue_date = {January 2025},
publisher = {VLDB Endowment},
volume = {18},
number = {5},
issn = {2150-8097},
url = {https://doi.org/10.14778/3718057.3718068},
doi = {10.14778/3718057.3718068},
journal = {Proc. VLDB Endow.},
month = aug,
pages = {1398–1411},
numpages = {14}
}

@inproceedings{DBLP:conf/csl/BaganDG07,
  author       = {Guillaume Bagan and
                  Arnaud Durand and
                  Etienne Grandjean},
  editor       = {Jacques Duparc and
                  Thomas A. Henzinger},
  title        = {On Acyclic Conjunctive Queries and Constant Delay Enumeration},
  booktitle    = {Computer Science Logic, 21st International Workshop, {CSL} 2007, 16th
                  Annual Conference of the EACSL, Lausanne, Switzerland, September 11-15,
                  2007, Proceedings},
  series       = {Lecture Notes in Computer Science},
  volume       = {4646},
  pages        = {208--222},
  publisher    = {Springer},
  year         = {2007},
  url          = {https://doi.org/10.1007/978-3-540-74915-8\_18},
  doi          = {10.1007/978-3-540-74915-8\_18},
  bibsource    = {dblp computer science bibliography, https://dblp.org}
}

@inproceedings{AJAR2016,
author = {Joglekar, Manas R. and Puttagunta, Rohan and R\'{e}, Christopher},
title = {AJAR: Aggregations and Joins over Annotated Relations},
year = {2016},
isbn = {9781450341912},
publisher = {Association for Computing Machinery},
address = {New York, NY, USA},
url = {https://doi.org/10.1145/2902251.2902293},
doi = {10.1145/2902251.2902293},
booktitle = {Proceedings of the 35th ACM SIGMOD-SIGACT-SIGAI Symposium on Principles of Database Systems},
pages = {91–106},
numpages = {16},
location = {San Francisco, California, USA},
series = {PODS '16}
}

\end{document}